\documentclass{WileyASNA-v2}
\articletype{Article Type}%
\def\I{{\sc i}}
\def\II{{\sc ii}}

\received{28 October 2019}
\revised{28 November 2019}
\accepted{10 December 2019}

\begin{document}

\title{Differential abundance analysis of Procyon and Theta Sculptoris:
Comparison with abundance patterns of solar-like pairs}

\author[1]{C. R. Cowley*}

\author[2]{K. Y\"{u}ce}

\author[3]{D. J. Bord}

\authormark{Cowley \textsc{et al}}

\address[1]{\orgdiv{Department of Astronomy}, \orgname{University of Michigan, Ann Arbor,
1085 S. University}, 
\orgaddress{\state{Michigan 48109-1107}, \country{USA}}}

\address[2]{\orgdiv{Department of Astronomy and Space Sciences, Faculty of Science}, 
\orgname{University of Ankara}, \orgaddress{\state{Ankara, TR-06100}, \country{Turkey}}}

\address[3]{\orgdiv{Department of Natural Sciences}, 
\orgname{University of Michigan--Dearborn}, 
\orgaddress{\state{4901 Evergreen Road, Dearborn, MI 48128}, \country{USA}}}

\corres{*C. R. Cowley \email{cowley@umich.edu}}


\abstract{The precision differential abundance (PDA) technique is applied to the mid-F stars
Procyon and $\theta$ Scl using spectra from the ESO UVESPOP library.
We relate PDA patterns to endogenous processes related to condensation or
to exogenous processes connected to Galactic chemical evolution (GCE).
We employ one-dimensional LTE models, but emphasize the use of 
weaker lines ($\leq$ 20 m\AA) than are typically used in such studies.
We compare our results with PDAs of solar-type stars.
Abundances and PDAs are determined for 28 elements: C, N, O, Na, Mg,
Al, Si, S, Ca, Sc, Ti, V, Cr, Mn, Fe, Co, Ni, Cu, Zn, Y, Zr, Ba, La, Ce,
Nd, Sm, Eu, and Gd.
A plot of PDAs ($\theta$ Scl minus Procyon) vs. $Z$ shows a highly
significant correlation.  Moreover, local substructure of the plot
for the elements Ca-Zn and neutron-addition elements
is similar to that which can be found for solar twins.
Our PDA vs. $Z$ plot structural similarity 
to plots that can be made from the extensive work of Bedell et al. (2018).
That PDA structure and substructure is clearly a function of age.
}

\keywords{stars:abundances, stars:solar-type, stars:individual: Procyon, 
stars:individual:$\theta$ Scl}



\maketitle



Lawrence Aller (1963) discussed the differential abundance method
used by mid-20$^{th}$ century astronomers.  In it, one characterized 
stellar atmospheres by single, mean values of temperature and pressure.
The method allowed one to 
avoid the use of $gf$-values which were often highly uncertain.
This 
technique has been revised and updated in connection with analyses of
solar twins (Melendez, et al. 2009; Nissen \& Gustafsson 2018).  In the modern technique,
one determines basic stellar parameters and abundances from individual lines
in the standard  way (Gray 2005), using model atmospheres, 
equivalent widths and/or spectral synthesis.  
Differential abundances are then calculated from corresponding line pairs
in the twin stars.  

A primary advantage of the
historical work, near independence from $gf$-values, is retained.
The stars are chosen to have identical or nearly identical spectral types,
and presumably closely comparable structures and abundances--to be ``twins."  
The precision of the method 
has revealed significant differences which may arise as a result of 
star formation and Galactic chemical evolution (Bedell et al. 2018,
henceforth, BD18).
 
\section{The stars and their spectra \label{sec:stsp}}

Ramirez, Mel\'{e}ndez \& Asplund (2014, henceforth RAM14) carried out an 
extensive study of 
mid and late F-star pairs.
They noted that the thinness of the F-star convection zones relative to
those of the well-explored solar twins would make it easier to detect
abundance anomalies due, for example, to planet formation.  
Rather than making comparisons with the Sun,
they defined standard stars from within their sample.    
 
In the current work, we choose to compare two similar mid-F stars,
slightly hotter than those of the RAM14 sample. 
Procyon (HD 61421) is a well-studied binary, consisting of an F5 IV-V and a white
dwarf.  Its composition is generally considered
``indistinguishable from solar" (Liebert et al. 2013).  The star, 
$\theta$ Scl (HD 739),
is a single-lined binary (Fuhrmann \& Chini 2012, Fuhrmann et al. 2017).
Its spectral type is F5 V.  Contributions from the companions of either
star are neglected in this study, although Procyon could be affected by
pollution from its companion.  

Numerous studies have given abundances for Procyon
(Liebert et al. 2013).
Selective abundances for $\theta$ Scl  
are given by Fuhrmann et al. (2017) and
Battistini \& Bensby (2015).  The former work is not differential,
while latter compares F-stars
with the Sun and therefore is distinct from twin studies, which usually
involve stars with effective temperatures differing by some 100K or less.
 
Procyon and $\theta$ Scl are in the ESO UVESPOP library 
(Bagnulo, et al. 2003)  
which provides the high-quality
spectra suitable to apply the 
differential-twin abundance method in the F-star domain.  The
spectra are of comparable quality to those used used to analyze solar
twins (S/N 300-700, RP 80,000).  

\section{Selection and measurement of the line spectra \label{sec:sar}}

Lines were chosen from lists from Bedell et al. (2014)
and Nissen (2015), but largely  
from the Procyon identification list 
published on Cowley's web site 
\footnote{http://www-personal.umich.edu/$\sim$cowley/procyon/wlid.html}. We 
selected lines with no obvious blends for which the measured and Ritz 
wavelengths were $\le 0.03$\AA\, apart. Emphasis was given to 
weak ($\le$ 20 m\AA) lines (see Sec.~\ref{sec:lls}). 
Each line was then synthesized to discern 
the contribution of possible blends. The decision to use a particular 
line was aided by a synthesis with most of the broadening turned off, 
e.g. no turbulent or instrumental broadening. Additional useful 
information came from a synthesis with the contribution from the line 
in question turned off.  

We chose to measure equivalent widths with the automatic routine ARES 
(Sousa et al. 2015).  The code was downloaded from S. G. Sousa's web site
\footnote{http://www.astro.up.pt/$\sim$sousasag/ares/} and installed under LINUX. 
Our own measurements, especially those of KY, using IRAF's Gaussian 
fits, agreed well with those of ARES.  
We decided that automatic measurements
would remove any influence of a personal equation from the results if
the input parameters to ARES are the same for both stars.

\section{Models and software \label{sec:modsoft}}

Atlas9 models with solar abundances were used by both CRC and KY.
Solar abundances are from
Asplund et al. (2009) with updates by Scott et al. (2015), and 
Grevesse et al. (2015).   
To obtain abundances from an equivalent width, KY used 
WIDTH9, while CRC fit Voigt profiles (cf. Cowley et al. 2014, and references
therein).  We used a set of 22 Fe {\sc i} and 33 Fe {\sc ii} lines.
The equivalent widths of these iron lines ranged from 3 to 90 m\AA. 
The mean difference of our measurements
of $\log(W)$, CRC$-$KY, was 0.0031 dex. These lines were used
to determine the microturbulent velocity and to ensure that the abundances were
independent of excitation potential.

The standard method of deriving the 
microturbulence is to require that the stronger lines agree in 
abundance with the weakest ones. However, this removes virtually all 
abundance information from the strongest lines. Because of this, we 
have based our final model parameters on  
12 Fe \I\, lines and 13 Fe \II\, lines 
with equivalent widths $\lessapprox 20 $m\AA~(Tab. 1).
Our calculations used LTE and 1-dimensional models as
is still common in differential analyses 
(Nissen \& Gustafsson 2018).

In order to choose values of $T_{\rm eff}$ and $\log(g)$ for abundance calculations,
four or five models were calculated, based on initial
estimates from the literature and spectral types.  Fe \I\, and \II\, abundances
($Ab$) were calculated for these models.
They were assumed to obey a linear relation:
\begin{equation}
Ab = a\cdot T_{\rm eff} + b\cdot \log(g) + c.
\label{eq:abc}
\end{equation}
\noindent The coefficients were determined separately by least squares for the
reduced set of Fe \I\, and Fe \II\, lines (cf. Tab. 1) and then
the right-hand sides of Eq.~\ref{eq:abc} were set equal to each other.
This gives a linear relation between $T_{\rm eff}$ and $\log(g)$, or a line on a
Kiel Diagram (Hunger 1955). 
This relation is also a function of the lines chosen and their 
$gf$-values (see Sec.~\ref{sec:lls}).

To fix our final $T_{\rm eff}$ and $\log(g)$, we used  Str\"{o}mgren 
(Smalley 2014) and Geneva (K\"{u}nzli et al. 1997) 
photometry, synthesis and comparison of the Balmer-line profiles, as well as 
the requirement that Fe \I\, and Fe \II\, yield the same abundance.  
Our adopted model for Procyon has $T_{\rm eff} = 6550$K
and $\log(g) = 3.9$, with $v\cdot\sin(i) = 3.0$ km/sec and a
microturbulence of 2.0 km/sec.  For $\theta$ Scl, these values
were 6525K, 4.3, 3.0 km/sec, and 1.7 km/sec.
A possible contribution from macroturbulence is included in our values
of $v\cdot\sin(i)$, which were found by spectrum synthesis.  
These figures are in agreement with values in the literature (cf.
Allende Prieto et al. 2002; Fuhrmann \& Chini 2012).
Uncertainties
in the models and their effect on our results are discussed in 
Sect.~\ref{sec:err}. 

\section{Lines and transition probabilities: abundances}
\label{sec:lls}
\subsection{Iron\label{ssec:iron}}

\begin{table}
\caption{Adopted weak Fe \I\, and Fe \II\, lines \label{tab:fe12}}
\centering
\begingroup
\setlength{\tabcolsep}{5pt}
\begin{tabular}{l l l|l l l} 
\hline\hline
\multicolumn{3}{c}{Fe I}&\multicolumn{3}{c}{Fe II} \\ \hline
$\lambda$(\AA)&$\log(gf)$& Desc.&$\lambda$(\AA)&$\log(gf)$& Desc. \\ \hline
4808.148 & -2.668 &    sol &    4720.139 &   -4.570 &   sol \\
5295.312 & -1.557 &    sol &    4833.192 &   -4.621 &   sol \\
5386.333 & -1.730 &    sol &    4953.980 &   -2.680 &   sol \\
5441.339 & -1.588 &    sol &    5000.730 &   -4.733 &   sol \\
5491.832 & -2.188 &    C+  &    5161.175 &   -4.203 &   sol \\
5696.089 & -1.720 &    C+  &    5427.816 &   -1.545 &   sol \\
5855.076 & -1.478 &    C+  &    5525.117 &   -3.970 &   C   \\
6226.734 & -2.098 &    sol &    5591.360 &   -4.527 &   sol \\
6271.278 & -2.703 &    C+  &    6113.319 &   -4.078 &   sol \\
6725.356 & -2.212 &    sol &    6179.390 &   -2.797 &   sol \\ 
6733.150 & -1.456 &    sol &    6239.357 &   -4.812 &   sol \\
6837.006 & -1.687 &    B   &    6248.907 &   -2.427 &   sol \\
         &        &        &    6446.407 &   -1.960  & V3 \\ \hline
\end{tabular}
\endgroup
\end{table}

The use of very weak lines has advantages and disadvantages.  Weak
lines cannot be measured as accurately as stronger ones.  In our experience,
this is true both for lines measured ``by eye'' as well at those measured
by ARES.  Additionally, reliable oscillator strengths are more difficult
to find for weak features (see below).  On the other hand, curve of growth effects 
(e.g., microturbulence and hyperfine structure) are
reduced by the use of weak lines.  In Cowley \& Y\"{u}ce (2019), we 
found that for a
typical Fe \I\, line with an equivalent width of 15.2 m\AA, the saturation
effect weakened the line by 17\%.  Although with differential methods,
oscillator strengths cancel, this is only true for very weak lines and
for lines where the saturation effects associated with turbulence and damping are
taken precisely into account. 

Oscillator strengths with NIST (Kramida et al. 2019) ratings of C  or better were available 
for a few of our weak lines. Results using values from BRASS (Laverick et al. 2019) 
or VALD3 (V3, Ryabchikova et al. 2015) for the remaining lines led to results with 
wide scatter.  For those lines, we  
ultimately chose solar $gf$-values derived in 
this work from ARES measurements using the Kurucz solar flux atlas 
(Kurucz et al. 1984).  The Fe lines and adopted $\log(gf)$ values are shown in
Tab. 1.  The column labeled ``Desc''  gives the
NIST accuracy rating or ``sol,'' to indicate that solar $gf$-values
were adopted.  Of 12 the Fe \I\, lines in Tab. 1, nine overlap
with Bedell et al (2014).  For these lines the agreement is excellent, especially for five of
the lines for which solar values were derived.  None of our Fe \II\,
lines overlapped with Bedell et al (2014).

\begin{table*}
\caption{Individual results \label{tab:results}}
\centering
\begin{tabular}{lr|rr|rr|rrr} 
\hline\hline
\multicolumn{2}{c|}{}&\multicolumn{2}{c|}{$\theta$ Scl}&\multicolumn{2}{c|}{Procyon}&\multicolumn{2}{c}{Difference} \\
Spec. & N & Ab & Err & Ab & Err& Val.& Err & Ref. \\   \hline
C I \rule{0pt}{16pt} 
      &  7 & -3.608  & 0.087  & -3.528  & 0.052  & -0.080  & 0.041  &1,3\\  
N I   &  3 & -4.204  & 0.047  & -3.982  & 0.078  & -0.222  & 0.036  &1 \\     
O I   &  4 & -3.195  & 0.060  & -3.201  & 0.056  &  0.006  & 0.067  &1 \\  
Na I  &  4 & -5.915  & 0.064  & -5.771  & 0.059  & -0.144  & 0.036  & 1\\  
Mg I  &  3 & -4.557  & 0.130  & -4.563  & 0.090  &  0.006  & 0.055  & 1,2\\  
Al I  &  2 & -5.726  & 0.160  & -5.616  & 0.134  & -0.110  & 0.026  & 1,2,3\\  
Si I  &  9 & -4.490  & 0.069  & -4.414  & 0.080  & -0.076  & 0.027  & 1,2,3\\  
S I   &  3 & -4.980  & 0.013  & -4.947  & 0.022  & -0.033  & 0.017  & 5\\  
Ca I  &  3 & -5.820  & 0.024  & -5.909  & 0.019  &  0.089  & 0.036  & 1,2\\  
Sc II &  3 & -9.107  & 0.067  & -9.051  & 0.074  & -0.056  & 0.052  & 1,20\\  
Ti I  & 13 & -7.176  & 0.027  & -7.155  & 0.043  & -0.021  & 0.019  & 1,6\\  
Ti II &  9 & -7.126  & 0.103  & -7.163  & 0.094  &  0.037  & 0.020  & 1,2,7\\  
V I   &  7 & -8.297  & 0.029  & -8.227  & 0.017  & -0.070  & 0.021   & 1\\  
V II  &  2 & -8.368  & 0.007  & -8.395  & 0.000  &  0.027  & 0.007   & 10,11\\  
Cr I  & 11 & -6.567  & 0.030  & -6.576  & 0.036  &  0.009  & 0.015  & 1,2,21\\  
Cr II &  2 & -6.427  & 0.119  & -6.420  & 0.070  & -0.007  & 0.050  & 8 \\  
Mn I  &  6 & -6.732  & 0.074  & -6.724  & 0.090  & -0.008  & 0.017  & 1,2\\  
Fe I  & 12 & -4.699  & 0.020  & -4.673  & 0.018  & -0.026  & 0.010  & 1,4 \\  
Fe II & 13 & -4.691  & 0.034  & -4.675  & 0.036  & -0.016  & 0.008  & 1,4\\  
Co I  &  6 & -7.068  & 0.154  & -7.026  & 0.141  & -0.042  & 0.014  & 1,2 \\  
Ni I  & 21 & -5.961  & 0.020  & -5.918  & 0.020  & -0.043  & 0.013  & 1,9\\  
Cu I  &  2 & -8.130  & 0.019  & -8.016  & 0.022  & -0.114  & 0.004  & 1,2\\  
Zn I  &  1 & -7.453  &\rule{0pt}{1pt}& -7.427  &\rule{0pt}{1pt}& -0.026  &\rule{0pt}{1pt}& 1  \\  
Y II  &  6 & -9.904  & 0.054  & -9.914  & 0.072  &  0.010  & 0.021   &1,12\\  
Zr II &  2 & -9.370  & 0.040  & -9.476  & 0.016  &  0.106  & 0.055   & 13,14\\  
Ba II &  2 & -9.899  & 0.045  &-10.086  & 0.040  &  0.187  & 0.005   & 1\\  
La II &  4 &-10.993  & 0.124  &-11.207  & 0.121  &  0.214  & 0.008   & 16\\  
Ce II &  3 &-10.402  & 0.056  &-10.510  & 0.063  &  0.108  & 0.075   & 15\\  
Nd II &  5 &-10.457  &-0.065  &-10.654  & 0.080  &  0.188  & 0.027   & 17\\  
Sm II &  3 &-11.071  & 0.054  &-11.226  & 0.087  &  0.155  & 0.035   & 18\\  
Eu II &  1 &-11.521  &\rule{0pt}{1pt}&-11.547  &\rule{0pt}{1pt}&0.026\rule{0pt}{1pt}&1,2 \\ 
Gd II &  2 &-10.482  & 0.126  &-10.610  & 0.190  &  0.127  & 0.065   & 19  \\ \hline 
\end{tabular}
\begin{tablenotes}{1 NIST, Kramida et al (2019); 2 Ryabchikova et al. (2015); 
3 Laverick et al. (2019); 4 solar gf;  
5 Zatsarinny \& Bartschat (2006); 6 Lawler et al. (2013); 7 Wood et al. (2013); 
8 Bouazza et al. (2018); 9 Ruczkowski, Elantkowska, \& Dembezy\'{n}ski (2017);
 10 Biemont et al. (1989); 11 Brewer et al (2016); 12 Palmeri et al. (2017); 
 13 Ljung et al. (2006); 14 Quinet, Bouazza \& Palmeri (2015); 
 15 Lawler et al. (2009); 16 Lawler et al. (2001)) 17 Den Hartog et al. (2003);
 18 Lawler et al. (2006); 19 Den Hartog et al. (2006); 20 Lawler \& Dakin (1989);
 21 Sobeck, Lawler \& Sneden 2007)}
\end{tablenotes}
\end{table*} 

\subsection{Other elements\label{ssec:other}}

\begin{table}
\caption{Sample data table for individual analyzed lines\label{tab:kytab}}
\centering
\begingroup
\setlength{\tabcolsep}{5pt}
\begin{tabular}{l  c  r r c c} 
\hline\hline
\rule{0.1mm}{0.0mm} &$\lambda$    &$W_{\lambda}$   &$\log$  &$\chi$  &$\log $      \\
Spectrum            & \AA         &  m\AA          & $(gf)$ & eV     &$(El/N_{\rm tot})$\\  \hline
C I                 &4817.373     & 8.7            &-3.080  & 7.480  & -3.317    \\
C I	                &5023.841     &15.8            & -2.210 &  7.950 &  -3.531   \\
C I	                &5551.579     & 7.0            &-1.900  & 8.640  & -3.690    \\ 
...                 &...          &...             &...     &...     &...        \\
Na I                &4497.657     &19.0            &-1.574  & 2.100  & -5.773   \\
Na I                &4751.822     &9.6             &-2.078  &2.100   &-5.617   \\
Na I                &5148.838     &5.6             &-2.044  &2.100   &-5.905    \\
...                 &...          &...             &...     &...     &...       \\  \hline
\end{tabular}
\endgroup
\begin{tablenotes}
{A machine-readable version of the complete table is available at the CDS.}
\end{tablenotes}  
\end{table}  

In order to select lines for analysis, 
ARES was given a list of all lines from the Procyon identification
list which had a single attribute whose wavelength differed by  
$\le 0.03$\AA.  We then chose lines, with a few exceptions,  having 
equivalent widths $\le 20$ m\AA.  Of the 174 lines used for abundances,
eight have equivalent widths $\ge 20$ m\AA.  The largest of these was
Eu \II\, with 45.5 m\AA\, in Procyon.  Generally, hyperfine structure
was not taken into account for our weak lines, but for this line in
Eu \II\ it was approximately taken into account using the data of
VALD3.  

The regions near all of these
lines were synthesized, as described for the iron lines.  While 
virtually all of the lines had small blends, lines were retained
unless the blends were estimated to contribute 10\% or more to  
equivalent widths.

All of the lines judged suitable for analysis in the Procyon spectrum
were adopted for $\theta$ Scl, provided there was an ARES measurement.
In a few cases, lines measured by ARES for Procyon were not measured
in $\theta$ Scl.  Usually such lines were discarded from the analysis,
but rarely, the lines were remeasured by CRC. 
Such measurements intercompare very well for weak lines with KY's
IRAF values.  The lines in question
are: Nd II 5092.77, 5092.81 and Gd II 4316.04.

We consulted the bibliographic reference data of 
NIST (Kramida et al. 2019) to select the most
recently determined oscillator strengths.  Where practical, we cite
original work rather than compilations (see Column 9 in Tab. 2).
 
Derived abundances are given in Tab. 2.  The abundances are
for $\log(El/N_{\rm tot}) = \log(El/H) - 0.036$.  
The Difference column is for ($\theta$ Scl minus
Procyon).  The error of the differences is calculated for the N individual
line pairs for each spectrum.
Data for individual lines are given in Tab. 3.   

\section{Errors}
\label{sec:err}
We have listed standard errors in Tab. 2, which are smaller
by $\sqrt{N}$ than standard deviations, which can readily be obtained using
Tab. 2.  In general, while standard errors are
technically correct, standard deviations often give a more realistic estimate of the
uncertainties because statistical analysis does not account for systematic errors.
The present differential technique should eliminate many systematic errors, however. 

The average standard error of  
our differential abundances (Column 8 of Tab. 2) is 0.0294 dex.
Overall, we have not reached the level of accuracy, 0.01 dex, of 
the solar twin studies (e.g. Mel\'{e}ndez et al. 2009; Nissen 2015), although for a few species,
including Fe \I\, and \II\,, our standard errors are of the order of 0.01 dex.

Uncertainties in the model parameters introduce some error.
Cowley \& Y\"{u}ce (2019, see Series 3 slide 3) 
gave results of a differential analysis for two sets of parameters
describing Procyon and $\theta$ Scl.  One set of results for Fe \I\, and Fe \II\
used the current parameters ($T_{\rm eff}, \log{g}$), 
the other used (6550K, 3.71) for Procyon and
(6550, 4.0) for $\theta$ Scl.  The differential abundances
for Fe \I\, and Fe \II\, were only 0.011 and 0.004 dex, respectively.

To test the sensitivity of the method to the assumed effective temperature, we
retain the parameters $T_e$ and $\log(g)$ for Procyon, and use a temperature 
100K lower for $\theta$ Scl: $T_e = 6425K$.  To ensure equal abundances for 
Fe \I\, and \II\, we use $\log(g) = 4.16$.  The resulting differences,
$\theta$ Scl minus Procyon then change from the values of Tab. 2, to -0.078
for Fe \I\. and -0.063 for Fe \II.  The change in the differences is 
a significant 0.05 dex.

NLTE corrections were available for a number of our lines from the web site 
at the Max Planck Institute \footnote{nlte.mpia.de} (Bergemann \& Nordlander 2014).  The 
differential corrections were generally quite small ($\le 0.01$ dex).
For Ti \I\, and Cr \I\,, average corrections
were +0.01 and +0.02 dex, respectively, discernable on a graph of our 
results (Fig. 1), but too small to affect the overall conclusions.

A more important source of error arises from the measurement of the equivalent widths.
ARES provides an error estimate, $\Delta W$, for each line (Sousa et al. 2015).   We used 
a sample of 194 Fe \I\, lines in Procyon between 3.06 and 20 m\AA.  When a line is partially
blended or has close neighbors, ARES will model the region, and it gives the number
of lines taken into account in determining the equivalent width of an individual
line.  We took cases where there were no or at most one blending line. 
Then, the average value
of $\Delta W/W$ is 0.0966.  We conclude that a typical
equivalent width could be $W \pm 0.0966W$, or roughly
$\log(W \pm \Delta W) = \log(W) \pm 0.04$.  We take this $\Delta W$ to be the standard 
deviation of the ARES $\log(W)$'s, and compare them with the standard deviations
of our logarithmic abundances.  The assumption here is that for weak lines,
the abundances are proportional to the equivalent widths. 

We thus compare this (0.04 dex) standard deviation with estimates of the standard deviation
of calculated abundances. 
We used variances in the differential abundances for 
12 Fe \I\, lines and 21 Ni \I\, lines  (Tab. 2), 
taking them to be representative.
Here the variances are $\sqrt{N}$ times the standard errors in the penultimate
column of Tab. 2.  We obtain 0.033 dex for the 12 Fe \I\, lines
and 0.063 for the 21 Ni \I\, lines.  These are of the same order as
the 0.04 dex we derived for the ARES measurements.  
This is the basis of our belief that the major source of uncertainty in our abundances
is the $\log(W)$ measurements. Examination of line-by-line fits and measurements by 
ARES leads us to conclude
that such errors arise as a result of both noise in the spectra and
differences in the normalization of the two spectra, and not from the
ARES technique itself.    
\section{Abundance trends\label{sec:trends}}

   \begin{figure}     
   \centering
   \includegraphics[angle=-0,width=8cm]{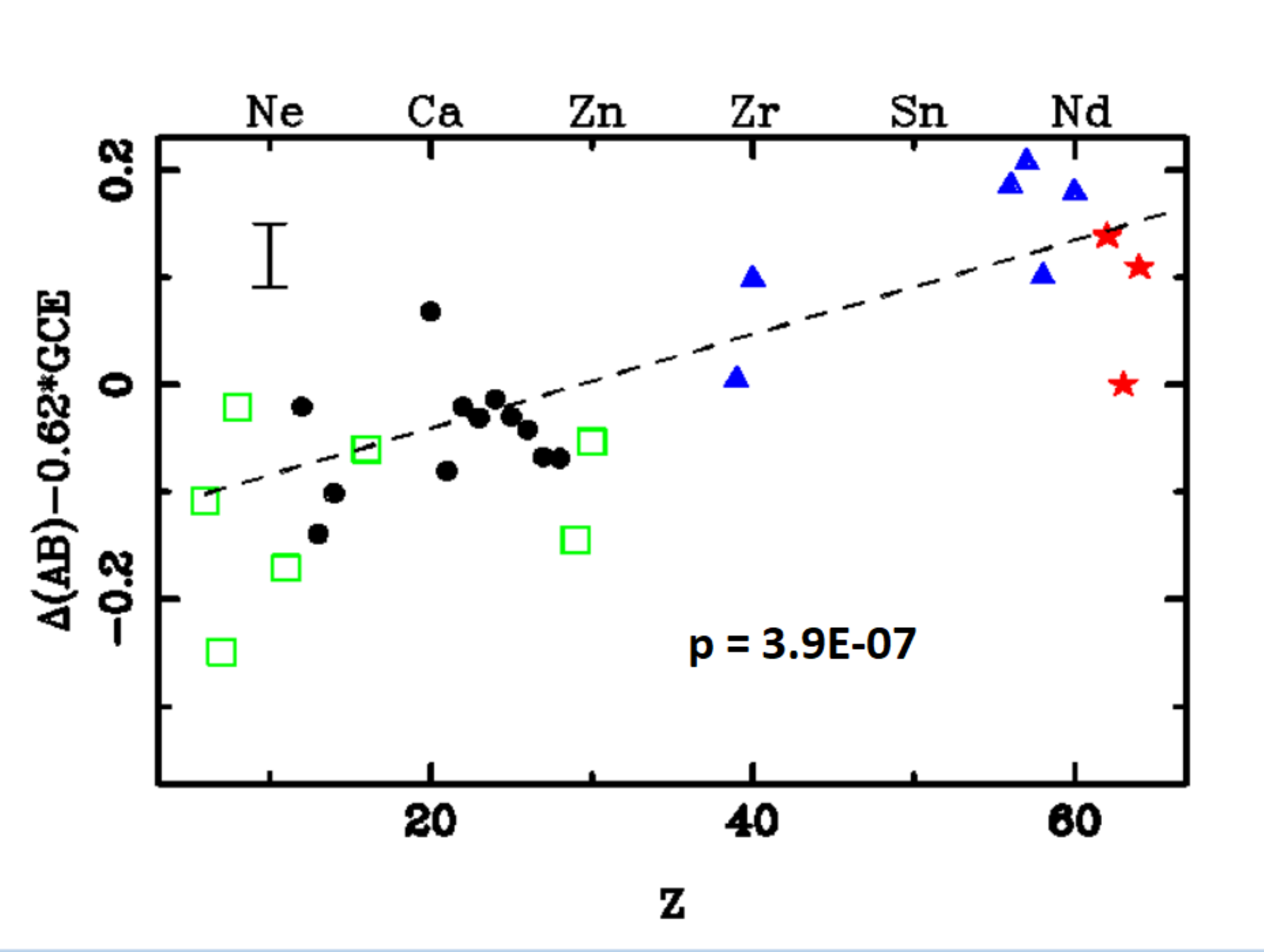}
      \caption{Differential logarithmic abundances ($\theta$ Scl minus Procyon) 
      vs. atomic number Z,
with corrections for GCE using an age difference of 0.62 Gyr.
The vertical bar has a length of twice the average standard error
(Col 8 of Tab.~\ref{tab:results}).  The dashed line is
$\Delta({\rm AB})-0.62\cdot {\rm GCE} = 4.38\cdot 10^{-3}Z - 0.128$.
Green squares are
volatile elements C, N, O, Na, S, Cu, and Zn.  Blue triangles are elements with
a dominant s-process contribution, while red stars have a dominant r-process
contribution. 
              }
         \label{fig:GCE}
   \end{figure}
   
When the differential abundances of $\theta$ Scl minus Procyon are plotted 
versus atomic number, a highly significant relation emerges.
We use Tab. 3 of BD18 to see if galactic chemical evolution (GCE)
could account for that correlation.  While that table is for the evolution of
[El/Fe], the correction to our [El/H] for a relatively short time interval
is negligible.  We take 0.62 Gyr as a representative age difference
based on the 2D linear interpolated values of
David \& Hillenbrand (2015) who give 1.48 and 2.10 Gyr for the ages of $\theta$ Scl and
Procyon, respectively. A slightly smaller age difference (0.30 Gyr) results
from using their 1D model (most probable) values, but differences in the range
of 0 to $\sim 4$ Gyr based on 68\% confidence limits cannot be ruled out. 

We subtract
from each differential abundance an amount $m\Delta t$, where $m$ is the slope of the
BD18 GCE relation for each element, and $\Delta t$ the difference in the ages of 
Procyon and $\theta$ Scl.     
The result is shown in
Fig. 1.  For convenience, we refer to the differential
abundance as $\Delta$(AB), rather than [El/H] since the latter is
traditionally used for abundance differences with the Sun.
Thus, $\Delta$(AB) = $[El/H]_{\theta {\rm Scl}} - [El/H]_{\rm Procyon}$  

A linear least-squares fit to the data of Fig. 1 is indicated by the
dashed line: $\Delta$(AB)$ = 4.38\cdot 10^{-3}Z -0.128$.  We shall refer to
the coefficient $4.38\cdot 10^{-3}$ as the ``overall slope'' to avoid confusion
with other local slopes to be mentioned below.
The fit has a Pearson
correlation coefficient of 0.7849 for 28 points.  The corresponding probability
that the relation arises by chance is $\sim 3.9\cdot 10^{-7}$.  
                                 
The overall slopes of plots like that of Fig. 1 for stars of the 
BD18 sample are well correlated with age (see Sec.~\ref{sec:bedplts}, Spina et al. 2018),
though there is considerable scatter.  Only the youngest stars in the BD18 sample
(0.5-0.6 Gyr) have slopes as large as $4.38\cdot 10^{-3}$, and in
this case the age difference of these stars with the Sun is some 4 Gyr.
It is unlikely that Fig. 1 for our stars, whose probable age 
difference is less than $\sim 1$ Gyr, has its slope solely as a result of GCE.
By the same token, given the uncertainties of the age estimates and the scatter
in the relation between slope and age, we cannot completely exclude some contributions
from GCE.

   \begin{figure}   
   \centering
   \includegraphics[angle=-0,width=8cm]{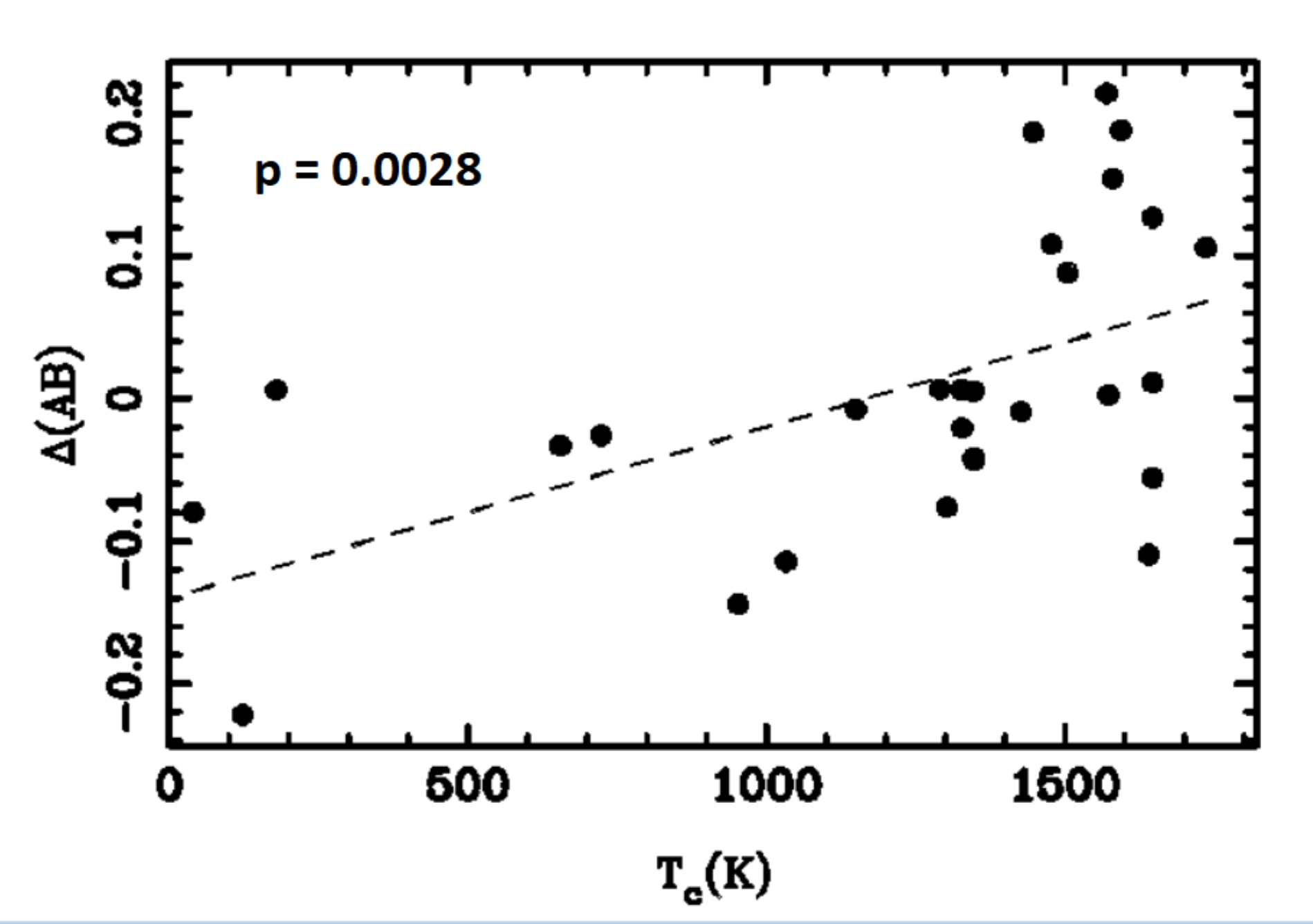}
      \caption{Differential abundances vs. Condensation temperature.
      The dashed line has the equation $\Delta({\rm Ab}) = 1.20\cdot 10^{-4}Z -0.140$.}
              
         \label{fig:vTc}
   \end{figure}  
                                 
In Fig. 2, we compare our differential abundances,
$\Delta({\rm Ab})$, with the 50\% condensation temperature, $T_c$ (Lodders 2003).
The small correction for GCE has not been made for this plot.
The probability that this plot arises by chance is 0.0028.
One can make a case, though not a strong one, for the relevance of
temperature-dependent condensation.  
Similar plots have been made by many authors in connection with solar twins
(Nissen \& Gustafsson 2018).   Much tighter correlations are
shown by Mel\'{e}ndez et al. (2009). 

Similar trends with $T_c$ are found in a wide variety of astronomical
sources from the interstellar medium (ISM, Jenkins 2003), $\lambda$ Boo stars 
(Heiter 2002), Post-AGB stars (Van Winckel 2003), Herbig Ae/Be stars
(Folsom et al. 2012), and solar twins (Mel\'{e}ndez 2009).  The first ionization potential 
effects (FIP, Laming 2015) observed in the solar and stellar coronae
separate elements such as C, N, and O from heavier, iron-group elements.
These chemical anomalies cover a wide range of magnitudes, from several
dex in the ISM to a few hundredths of a dex in the Sun vs. solar twins.
They also occur in a wide variety of astronomical settings.  It is not
surprising that the observed abundance patterns are variegated.  
                                 
Ample discussion exists in the literature
of possible scenarios that might explain such observations.  These range from
consequences of terrestrial planet formation to accretion of differentiated
interstellar material.   We refer to Nissen \& Gustafsson (2018, see Sec. 4), 
RAM14, and other cited papers for details.   

\section{Abundance patterns in the Bedell sample}                
\label{sec:bedplts}
\begin{figure}   
\centering                  
\includegraphics[angle=-0,width=8cm]{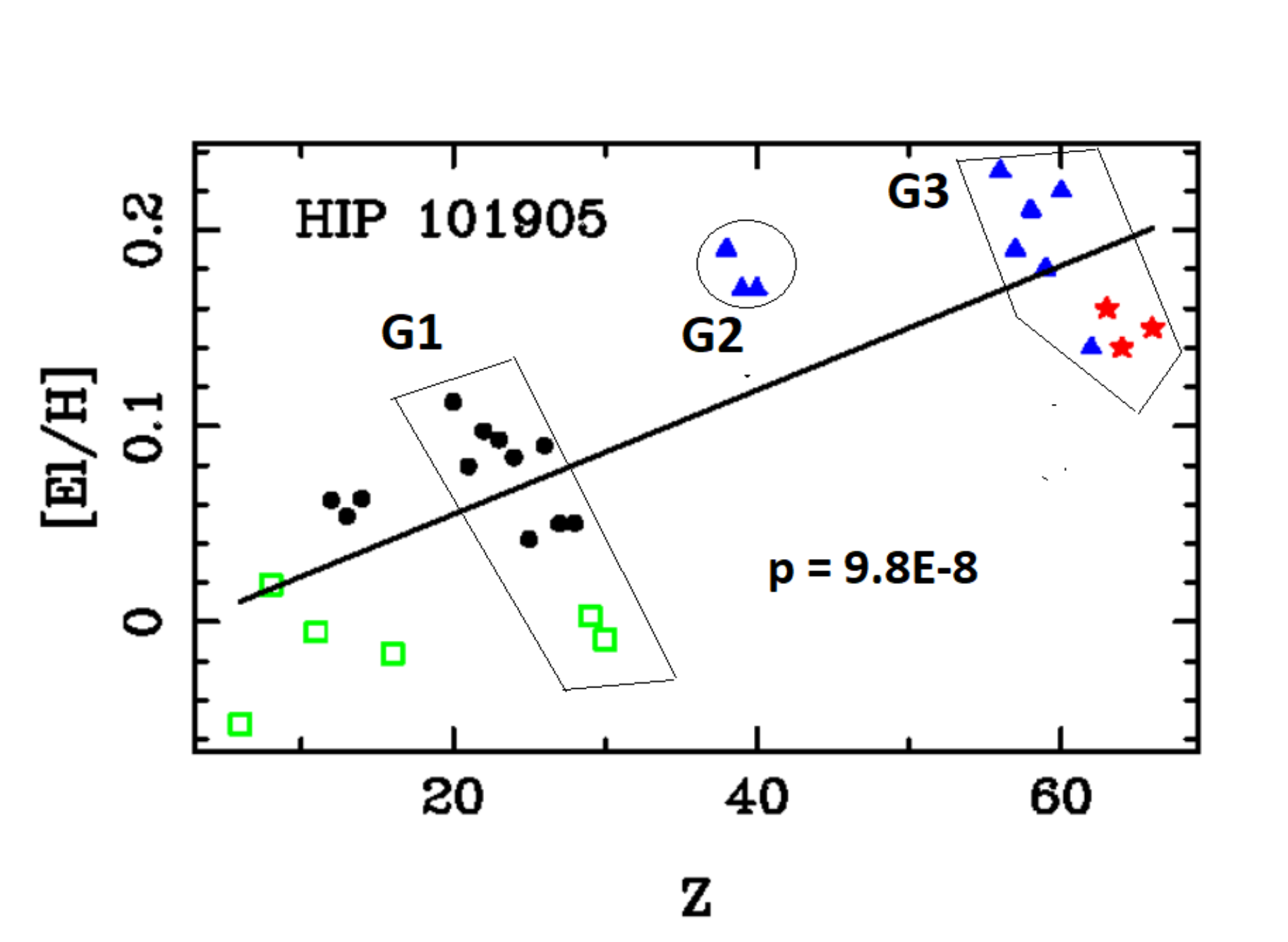}
\caption{Differential abundances for HIP 101905 vs. $Z$ showing characteristic
trends and groupings of a major portion of the BD18 stars.  
The age of HIP 101905 is estimated to be 1.2 Gyr.
The solid line has the equation: $[El/H]= 3.18\cdot 10^{-3}Z-0.0088$.
Symbols are as in Fig.~\ref{fig:GCE}. 
}
\label{fig:3G101905}
\end{figure} 

We had expected that the differential abundances of $\theta$ Scl vs. Procyon
to be random, as the stars were so similar in spectral type and
population.  The structure displayed in Fig. 1 led us to
examine other differential results.   
                                 
BD18 published precision differential 
abundance results for 79 solar-type stars.  The focus of that and larger studies
by Delgado Mena et al. (2019) or Brewer et al. (2016)
primarily on the element to element abundance variations.  However, the
coverage of $Z$ in the latter studies was not as complete as BD18,
so we do not discuss them further here.    

Our plots vs. $Z$ display the collective behavior of numerous elements.
We find a variety of patterns, many of which resemble our Fig. 1.
Fig. 3 is an example of one such plot for 30 elements, carbon
through dysprosium.  
Of BD18's 79-star sample, 35 stars show overall
linear fits that are significant at the $\le 0.01$ probability level.  
The overall slope of those fits correlate well with age.
The ``local'' groups G1 and G3 have significant negative slopes.
G1 consists of Ca ($Z$=20) through Zn ($Z$=30), of the first 
long period of the Periodic Table.  A straight-line fit with significance
$\le 0.01$ can be obtained for the G1 points for about a third of the BD18 
sample.    
G1 remains a coherent configuration
even in some stars where the overall slope is no longer well defined,
as is typical for many of the older stars.
Note that G1 consists of a sequence of elements with  decreasing $T_c$ and FIP.

\begin{figure}                   
\centering                  
\includegraphics[angle=-0,width=8cm]{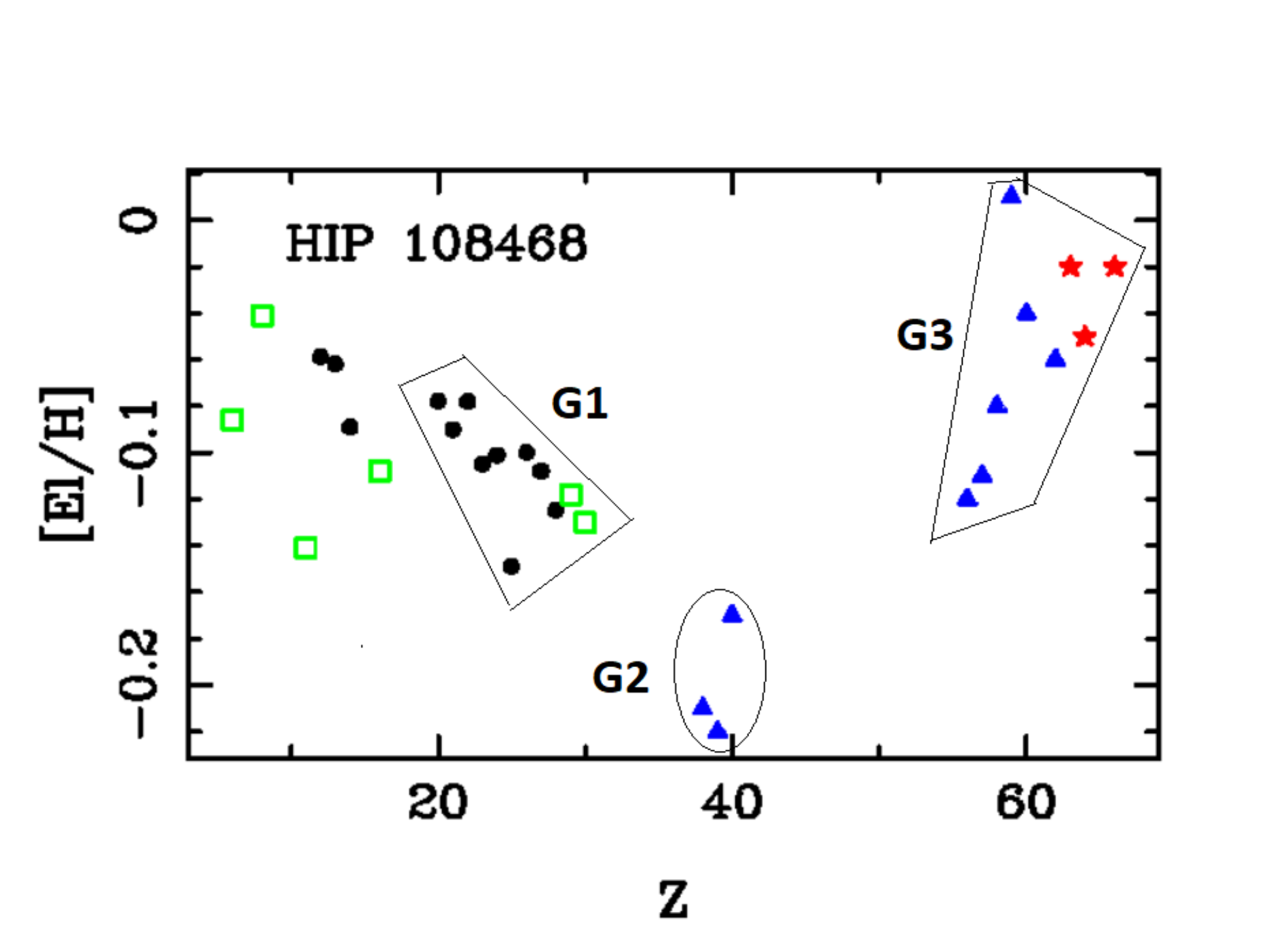}
\caption{Differential abundances for the 7.4 Gyr star HIP 108468 vs. $Z$,
illustrating the second major differential abundance pattern of the BD18 stars. 
Symbols are as in Fig.~\ref{fig:GCE}.
}
\label{fig:hip108468}
\end{figure}

Roughly a third of BD18's older stars have "V-like" shapes, due to the drop of 
the G2 group, as shown in Fig.~\ref{fig:hip108468}.
Note that G3 has rotated, and now has a positive slope. This positive
slope is characteristic of older stars, and contributes to the V-like pattern.
\section{Summary\label{sec:sum}}

The technique of precision differential abundances has been applied to the
closely similar F-stars Procyon and $\theta$ Scl.  A plot of abundance 
differences in the sense $\theta$ Scl minus Procyon against atomic number $Z$
shows a highly significant positive slope and a distinct non-random pattern.
This trend is unlikely to be due solely to Galactic chemical evolution.  
Similar patterns are found among solar-type stars in the survey of BD18, where
nearly half of the 79 stars included in that sample display statistically
significant overall fits, the (positive) slopes of which are correlated with
age.  Older stars of the BD18 collection often display markedly different
patterns of differential abundance as a function of $Z$.  A full description
of the patterns and correlations within the BD18 sample will be presented in
a forthcoming paper.  Differential abundances for $\theta$ Scl and Procyon
appear to be only weakly correlated with condensation temperature.

\section*{Acknowledgments}

This work made use of the VALD database, operated at Uppsala University, the
Institute of Astronomy RAS in Moscow, and the University of Vienna.  We
acknowledge with thanks the online facilities of NIST and the Belgian Repository
of Fundamental Atomic Data and Stellar Spectra.  We acknowledge use of data
from the UVES Paranal Observatory Project (ESO DDT Program ID 266 D-5655).
We thank Megan Bedell and Richard Monier for comments.  
CRC thanks Pierre North and      
Barry Smalley for photometric codes.  He also thanks his Michigan colleagues
and Kohei Hattori for
help and advice.  We thank the referee for useful comments and suggestions.
KY thanks Saul J. Adelman for help and encouragement.

\end{document}